# In-Plane Field induced Quantized Longitudinal Conductivity in Magnetic Topological Insulators

Ting-Hsun Yang[1,†], Yaochen Li[1,†], Peng Zhang[1], Penghao Zhu[2], Hung-Yu Yang[1], Eun Sang Choi[3], Kaiwei Chen[4], Wenqiang Cui[4], Kin Wong[1], Peng Deng[4], Gang Qiu[1,5,*], and Kang L. Wang[1,*]

[1]*Department of Electrical and Computer Engineering, University of California, Los Angeles, California 90095, United States*

[2]*Department of Physics, The Ohio State University, Columbus, Ohio 43210, United States*

[3]*National High Magnetic Field Laboratory, Florida State University, Tallahassee, Florida, 32310-3706, United States*

[4] *Beijing Academy of Quantum Information Sciences, Beijing, 100193, China*

[5]*Department of Electrical and Computer Engineering, University of Minnesota, Minneapolis, 55455, United States*

[†]These authors contribute equally to this work.

[*]Corresponding authors: gqiu@umn.edu, wang@ee.ucla.edu

**Abstract**

We report the discovery of an in-plane quantization (IPQ) state in trilayer magnetic topological insulators, characterized by a quantized longitudinal conductivity of $\sigma_{xx} = e^2/h$ under strong in-plane magnetic fields. This state emerges at a quantum critical point separating quantum anomalous Hall phases tuned by field angle and orientation, directly linking gap-closing behavior to quantized criticality. Temperature and gate dependent transport measurements, supported by a self-consistent approximation model, reveal that electron–hole puddles dominate charge transport in this regime, highlighting the essential role of impurity disorder in stabilizing quantized critical transport. These findings establish a tunable experimental framework that connects gap-closing physics with universal conductivity, offering both microscopic insight into critical transport in magnetic topological insulators and a robust platform for probing quantum criticality in topological systems.

**Introduction**

In magnetic topological insulators (MTIs), diverse quantum phases emerge depending on the magnetic configuration and material conditions. With uniform magnetization, the system becomes a Chern insulator, exhibiting a gapped surface state and chiral edge channels that lead to the quantum anomalous Hall effect (QAHE) [1–3]. Antiparallel surface magnetizations stabilize the axion insulator, characterized by a nontrivial axion angle and quantized magnetoelectric response [4–6]. Additional insulating regimes include the Anderson insulator, driven by disorder localization [7], and the normal insulator that appears near coercivity due to strong domain scattering [8,9] or band inversion [10,11]. At the boundaries between these localized states, the system undergoes a quantum critical point (QCP) where electronic states become extended. Remarkably, across multiple MTI systems, a nearly quantized longitudinal resistance of $h/e^2$ has been observed at these QCPs [7,11–13], indicating a universal, materials-independent critical behavior.

To better access this critical regime, novel scaling methods have been developed in MTIs [13], wherein an effective magnetic field is redefined through temperature-dependent coercivity. This approach enables precise tuning to the QCP, where quantized resistance and half-quantized Hall conductivity emerge, consistent with the phenomenological predictions of the Kivelson–Lee–Zhang (KLZ) model for Chern–Simons bosons at criticality [14]. In KLZ, duality and scaling symmetry constrain conductivities to universal values, but the model does not provide a microscopic mechanism for quantization. This limitation is underscored by the strong sample dependence of the critical point in MTIs, which shifts with film thickness [15,16] and dopant concentration [17]. This sensitivity complicates the universal observation of QCP signatures and highlights the need for new experimental platforms.

In this work, we report robust quantization of the longitudinal conductivity in trilayer MTIs under strong in-plane magnetic fields. By systematically varying the field direction and magnitude, we uncover a field-driven renormalization-group (RG) flow within a single material platform, revealing a continuous evolution of quantum critical behavior. This transition arises from exchange-gap closing and the emergence of electron–hole puddles that dominate transport near the Dirac point, as supported by gate-dependent measurements and a disorder-driven self-consistent theoretical model. Together, these results establish a tunable experimental framework for probing universal aspects of quantum criticality and provide evidence toward a microscopic understanding of critical transport in MTIs.

**Result**

The MTI heterostructures were grown by molecular beam epitaxy (MBE) and consist of a trilayer stack comprising 3 quintuple layers (QLs) of V-doped (Bi, Se)$_2$Te$_3$ (V-BST), 4 QLs of (Bi, Se)$_2$Te$_3$ (BST), and 3 QLs of Cr-doped (Bi, Se)$_2$Te$_3$ (Cr-BST). The films were patterned into Hall bar and Corbino disk geometry using photolithography and both DC and AC transport measurements were performed at 30 mK (see Supplementary note 1). To investigate the magnetic field driven behavior, we first examined the longitudinal resistance ($R_{xx}$) and Hall resistance ($R_{xy}$) of the Hall bar device under an in-plane magnetic field applied along the x-y plane, as shown in Fig. 1(a). Prior to the in-plane field sweep, the samples were initialized into the quantum anomalous Hall (QAH) state by applying a 2 T perpendicular field to saturate the magnetization along the out-of-plane (z) axis. Upon removal of the perpendicular field, the remanent magnetization sustains the QAH phase, characterized by a quantized $R_{xy} = h/e^2 \approx$ 25.8 k$\Omega$ and vanishing $R_{xx}$ shown at zero in-plane field. As the field increases, the magnetization gradually tilts toward the plane of the film. This tilting reduces the out-of-plane component necessary to preserve the exchange gap, leading to breakdown of chiral edge transport [11,18]. Consequently, $R_{xy}$ decreases from its quantized value while $R_{xx}$ rises due to the dissipative bulk conduction. At sufficiently large in-plane fields (e. g., 8T), both $R_{xx}$ and $R_{xy}$ reach saturation, reflecting full alignment of the magnetization within the plane. In the Fig. 1(b), the resistance data are converted into longitudinal ($\sigma_{xx}$) and Hall ($\sigma_{xy}$) conductivity using standard tensor relations:

$$\sigma_{xx} = \frac{\rho_{xx}}{\rho_{xx}^2 + \rho_{xy}^2}, \quad \sigma_{xy} = \frac{\rho_{xy}}{\rho_{xx}^2 + \rho_{xy}^2}. \tag{1}$$

where $\rho_{xx}$ and $\rho_{xy}$ denote the longitudinal and Hall resistivities, respectively, calculated using a device aspect ratio L/W = 2. In the low-field regime, the system exhibits a quantized Hall and vanishing longitudinal conductivity, consistent with QAH state. Notably, at high in-plane fields, this behavior reverses: the $\sigma_{xy}$ is suppressed toward zero, while the $\sigma_{xx}$ saturated at the quantized value of $e^2/h$. This distinct high-field transport regime is referred to as in-plane quantization (IPQ) in the discussion that follows. In Fig. 1(c), similar IPQ has been found in the device with Corbino disk structure. Unlike the Hall bar structure, the Corbino disk inherently suppresses edge channel contributions, isolating the bulk response [19,20]. Following the same perpendicular field training, the sample exhibits extremely high two-terminal resistance $R_{2T}$, indicative of an insulating bulk characteristic of the QAH state. As the in-plane field increases, $R_{2T}$ decreases sharply and eventually saturates near 5.7 kΩ. Here, the bulk conductivity is extracted from the Corbino geometry using the relation [21]:

$$\sigma_{2T} = \frac{1}{2\pi R_{2T}} \ln\left(\frac{R_2}{R_1}\right). \qquad (2)$$

where $R_1$ and $R_2$ denote the inner and outer radii of the disk, respectively. The resulting conductivity, plotted in Fig. 1(d), closely mirrors the behavior of $\sigma_{xx}$ obtained from the Hall bar measurements in Fig. 1(b). This correspondence confirms that the IPQ regime reflects intrinsic bulk conduction properties that are distinct from edge-dominated quantum state like quantum spin Hall effect [22].

The observation of IPQ requires precise alignment of the external magnetic field within the sample plane, as even small angular deviations can obscure the features. To systematically explore this sensitivity and capture the full angular evolution of the system, we performed angle-dependent magneto transport measurements. As shown in Fig. 2(a) (see Supplementary note 3 for $R_{xx}$ and $R_{xy}$), a fixed magnetic field of 10 T was rotated continuously from the out-of-plane (θ = 0°) through the in-plane direction (θ = 90°) to the opposite out-of-plane orientation (θ = 180°), as illustrated schematically in the inset. When the magnetic field is aligned near out-of-plane direction (θ = 0° or 180°), the system retains signatures of the QAH state. This quantized value persists across a broad angular range, reflecting strong perpendicular magnetic anisotropy (PMA) that stabilizes the out-of-plane magnetization. As the field angle approaches in-plane direction, $\sigma_{xx}$ sharply increases and peaks at $e^2/h$, while $\sigma_{xy}$ simultaneously drops to zero, consistent with the IPQ state observed in Fig. 1.

The angular dependence of the topological phase transition can be further visualized by mapping the transport evolution onto a renormalization group (RG) flow diagram, as shown in Fig. 2(b). The resulting trajectory traces a pronounced semicircle connecting two stable fixed points at $(\sigma_{xx}, \sigma_{xy}) = (-e^2/h, 0)$ and $(e^2/h, 0)$, corresponding to QAH states with opposite Chern numbers of −1 and +1, respectively. Notably, a critical point emerges at the midpoint of the arc, located at $(0, e^2/h)$, which corresponds to the IPQ state. This flow pattern is distinct from that observed in conventional MTI systems, represented by the dashed line in Fig. 2(b), where the flow consists of two smaller semicircular arcs intersecting at the fixed point (0, 0), typically associated with trivial or axion insulator states [4,5,9,15,23,24]. To examine the role of magnetic field strength, we map the evolution of the RG flow under varying fields in Fig. 2(c). At low fields (e.g., 1 T), the system exhibits clear hysteresis between forward (0° to 180°) and backward (180° to 0°) angle sweeps, reflecting the distinct PMA energies of the Cr- and V-doped layers. As the field strength increases, this asymmetry diminishes and the forward and backward trajectories gradually merge, indicating a suppression of the layer-dependent magnetization switching. Concomitantly, the RG flow undergoes a transition: the two smaller semicircular trajectories coalesce into a single large semicircle, and the initial dips indicated by red triangles in lower field evolve into IPQ at the critical point. This field-driven transformation contrasts with earlier reports, where such RG behavior and quantized longitudinal conductivity were observed only in thicker MTI samples near the coercive field [15,25]. In our case, the IPQ state emerges under strong in-plane fields, highlighting magnetic field strength and orientation as a key tuning parameter for access QCP.

To elucidate the microscopic origin of the IPQ state, we examine the temperature dependence of the longitudinal conductivity $\sigma_{xx}$ under a constant magnetic field of 8 T, while systematically varying the field angle from out-of-plane ($\theta = 0°$) to fully in-plane ($\theta = 90°$), as shown in Fig. 3(a) (see Supplementary note 4 for $R_{xx}$ and $R_{xy}$). In the QAH regime at $\theta = 0°$ (green curve), $\sigma_{xx}$ remains near zero up to approximately 200 mK. As temperature increases, thermally activated bulk carriers contribute to longitudinal conduction, introducing dissipative channels and destabilizing the QAH state. With increasing field angle ($\theta \approx 60° \sim 80°$), the growing in-plane component of the magnetic field enhances conductivity at base temperatures and increasing temperature results in a monotonic rise in $\sigma_{xx}$. As the angle approaches $\theta = 90°$, the IPQ state emerges and maintains its quantized value in the low-temperature regime, similar to the QAH state. This quantized plateau is clearly visible in the Arrhenius plot of Fig. 3(b), where the log scale of $\sigma_{xx}$ is plotted against inverse temperature $1/T$. At high $1/T$ range (low temperature), $\sigma_{xx}$ remains pinned at $e^2/h$, in contrast to the thermally activated behavior observed at smaller field angles. The temperature dependent $\sigma_{xx}$ can be modeled using an Arrhenius relation [26]:

$$\sigma_{xx} = A \exp\left(\frac{-E_a}{k_B T}\right) + \sigma_0. \qquad (3)$$

where $E_a$ is the activation energy, $k_B$ is the Boltzmann constant, A is a device-specific perfector, and $\sigma_0$ captures residual conductivity from non-activated channels. The extracted activation energies are summarized in the inset of Fig. 3(b). At $\theta = 0°$, the activation gap is approximately 80 µeV, in close agreement with previous reports [16,26]. As the angle increases, the gap decreases monotonically and vanishes near $\theta = 90°$, indicating that the in-plane magnetic field effectively closes the surface exchange gap. This result agrees with prior observations of giant resistance changes in MTIs attributed to gap-closing behavior under in-plane fields [18].

The strong connection between surface gap closure and the emergence of the IPQ state is further investigated through gate-dependent transport measurements. As shown in Fig. 4(a), we measure $\sigma_{xx}$ (see Supplementary note 5 for $R_{xx}$ and $R_{xy}$) as a function of gate voltage under various in-plane magnetic fields, following the same perpendicular field training protocol described in Fig. 1(a). At zero in-plane field, the system remains in the QAH regime (green curve), characterized by near-zero longitudinal conductivity over a finite gate voltage range. In this regime, the asymmetrical conductivity profile indicates that the material is intrinsically p-type. As the in-plane field increases, finite longitudinal conductivity emerges even at zero gate voltage, signaling a breakdown of the QAH phase. Applying gate voltage introduces additional carriers, leading to a rapid increase in $\sigma_{xx}$. At sufficiently high in-plane field, the expected IPQ state appears. Remarkably, this quantized plateau can be sustained over a narrow gate voltage window, and the conductivity exhibits only weak dependence on gate voltage, in contrast to the behavior observed at lower in-plane fields. The distinct gate voltage ranges of the QAH and IPQ states are further highlighted in Fig. 4(b), where the derivative of conductivity with respect to gate voltage is plotted. For the QAH state, half of the magnetization-induced exchange gap $\Delta V_{QAH}$ is estimated from the gate range extending from the charge neutrality point (CNP), identified by the minimum in the conductivity curve, to the lower bound of the gap, approximately at 9.5 V. Although the gate voltage window for the IPQ state, $\Delta V_{IPQ}$, is considerably narrower than that of the QAH state, the flat region in the derivative plot at $B_{//}$ =14 T suggests a plateau width of about 3 V. Notably, this entire IPQ window falls within the exchange gap determined from the QAH phase, indicating that the in-plane quantized transport arises in a regime where the surface gap is nearly closed but not yet completely overlapped by bulk states.

Together with gap-closing behavior, gate dependence provides key insights into the nature of the IPQ state. We propose that this behavior reflects Dirac-like physics, reminiscent of the minimum conductivity observed in graphene [27,28]. In graphene, conductivity near $4e^2/h$ occurs at the Dirac point due to the formation of electron and hole puddles caused by charge inhomogeneity [29]. These locale puddles allow finite conduction even in regions that would otherwise be insulating. By analogy, a similar mechanism may be applied to trilayer MTI system. Fig. 4(c) and 4(d) illustrate the energy landscape and associated magnetic domain configurations for both $B_{//}$ = 0 T and $B_{//}$ = 14 T. In the absence of in-plane field, the Dirac surface states are gapped due to magnetic ordering, and the Fermi level lies within this gap, enabling dissipationless edge transport in the QAH regime. Although energy fluctuations exist [30,31], they are insufficient to induce surface conduction. However, at high in-plane fields (e.g., 14 T), the exchange gap closes, restoring Dirac-like surface states. Disorder-induced energy fluctuations then shift portions of the Fermi level above or below the Dirac point across the sample, forming a network of electron and hole puddles. This spatially varying carrier distribution enables IPQ to be observed at high in-plane fields.

The charge puddle-dominated transport near the Dirac point can be well described by the self-consistent approximation (SCA) model [32]. In this framework, impurity-induced energy disorder leads to spatial fluctuations in the electrostatic potential, forming localized electron and hole puddles. These puddles create a screening effect that suppresses further disorder, resulting in a residual carrier density $n^*$ determined by the balance between disorder and screening. While $n^*$ provides finite carriers that enhance conductivity, the impurity density $n_{imp}$ introduces scattering. Their interplay yields a nearly constant longitudinal conductivity for per surface of MTIs:

$$\sigma_{xx} \sim C \left| \frac{n^*}{n_{imp}} \right| \frac{e^2}{h}. \tag{4}$$

Here, $C$ is a dimensionless constant determined by the Wigner–Seitz radius $r_s$. For $Bi_2Se_3/SiO_2$, $C$ is estimated to lie in the range 30 – 300 with given $r_s$ [33]. In our analysis, we take the average value $C$ = 150. Combining this with the experimentally observed IPQ state yields a characteristic ratio of $n^*/n_{imp} = 1/300$. In a two-dimensional Dirac system, the carrier density $n$ is related to the Fermi energy via:

$$n = \frac{g}{4\pi} \frac{E_F^2}{\hbar^2 v_F^2}. \tag{5}$$

where $\hbar$ is the reduced Planck constant, $v_F$ is the Fermi velocity, and $g$ = 2 accounts for one Dirac cone on each surface in the trilayer MTI heterostructure. Combining Eqs. (4) and (5), we obtain the characteristic Fermi energy fluctuation induced by impurity density:

$$E_F^{fluc} = \hbar v_F \sqrt{\frac{\pi n_{imp}}{150}}. \tag{6}$$

In MTIs , magnetic dopants thus play a dual role: they open the exchange gap required for the QAH effect and simultaneously introduce potential disorder that generates charge puddles. Scanning tunneling microscopy studies [30] have shown a near-linear relationship between the exchange gap $\Delta_{ex}$ and dopant density. For instance, a Cr impurities density of $1 \times 10^{13}$ cm$^{-2}$ yields $\Delta_{ex} \sim 10$ meV, corresponding to half the full gap. This same impurity density produces $E_F^{fluc} \sim 12.7$ meV, by using a $v_F = 2.9\ eV \cdot Å$ [30]. Since carrier density scales as $n \propto E_F^2$ from Eq. (5), and the gate voltage linearly modulates carrier density, it follows that $\Delta V_g \propto E_F^2$. By combining the estimated $E_F^{fluc}$ with the measured QAH gate plateau width $\Delta V_{QAH}$, we estimate the expected gate range for the IPQ plateau to be $\Delta V_{IPQ}^{SCA} \sim 15.2$ V. Although this estimate is larger than the experimentally observed gate width $\Delta V_{IPQ}$ in Fig. 4(b), the discrepancy likely arises from the fact that charge puddle-dominated transport does not produce a perfectly flat quantized plateau. Allowing for a ± 5% deviation from e$^2$/h, the observed plateau width aligns more closely with the model prediction. Finally, Eq. (6) highlights that a higher impurity density supports the emergence of IPQ state. This explains why quantized criticality is more consistently observed in doped MTIs [7, 11-13], whereas in cleaner systems like undoped Cd$_3$As$_2$, deviations up to 20% from quantization have been reported under strong in-plane fields [34].

**Conclusion**

In conclusion, our study identifies the IPQ state as a robust transport signature of quantum criticality in trilayer MTIs under strong in-plane magnetic fields. By combining angle-, temperature-, and gate-dependent transport measurements with insights from the SCA model, we demonstrate that the IPQ state emerges from exchange gap closing and the formation of disorder-induced charge puddles near the Dirac point. These results highlight the critical role of impurity-driven mechanisms in enabling quantized critical transport and establish the IPQ state as a universal marker of restored gapless Dirac behavior in MTIs. Moreover, the shared characteristics of this regime with other Dirac systems open new avenues for investigating unconventional transport phenomena in MTIs, including Dirac fluid behavior [35] and hydrodynamic electron flow [36].

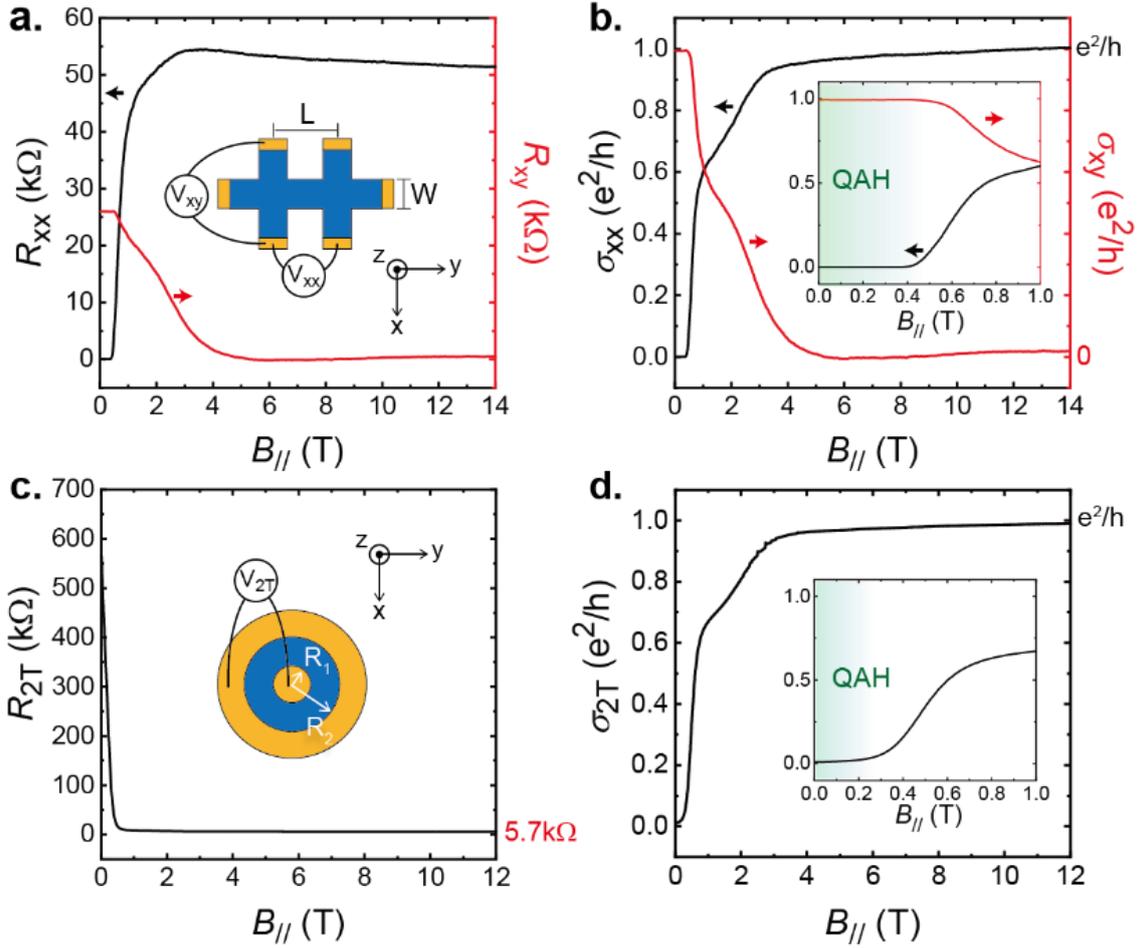

**FIG. 1.** (a) Longitudinal resistance ($R_{xx}$) and transverse resistance ($R_{xy}$), and (b) their corresponding conductivities ($\sigma_{xx}$ and $\sigma_{xy}$), plotted as a function of in-plane magnetic field after the sample's magnetization has been saturated using a 2T out-of-plane magnetic field in a Hall bar structure with a length-to-width ratio (L/W) of 2. The inset in panel (b) emphasizes the quantum anomalous Hall (QAH) state within a 1T in-plane field. (c) Two-terminal resistance ($R_{2T}$) and (d) conductivity ($\sigma_{2T}$) as functions of the in-plane field in a Corbino disk structure with a radius ratio of $R_1/R_2 = 1/4$, following the same out-of-plane field treatment. The inset in panel (d) also illustrates the QAH state within a 1T in-plane field.

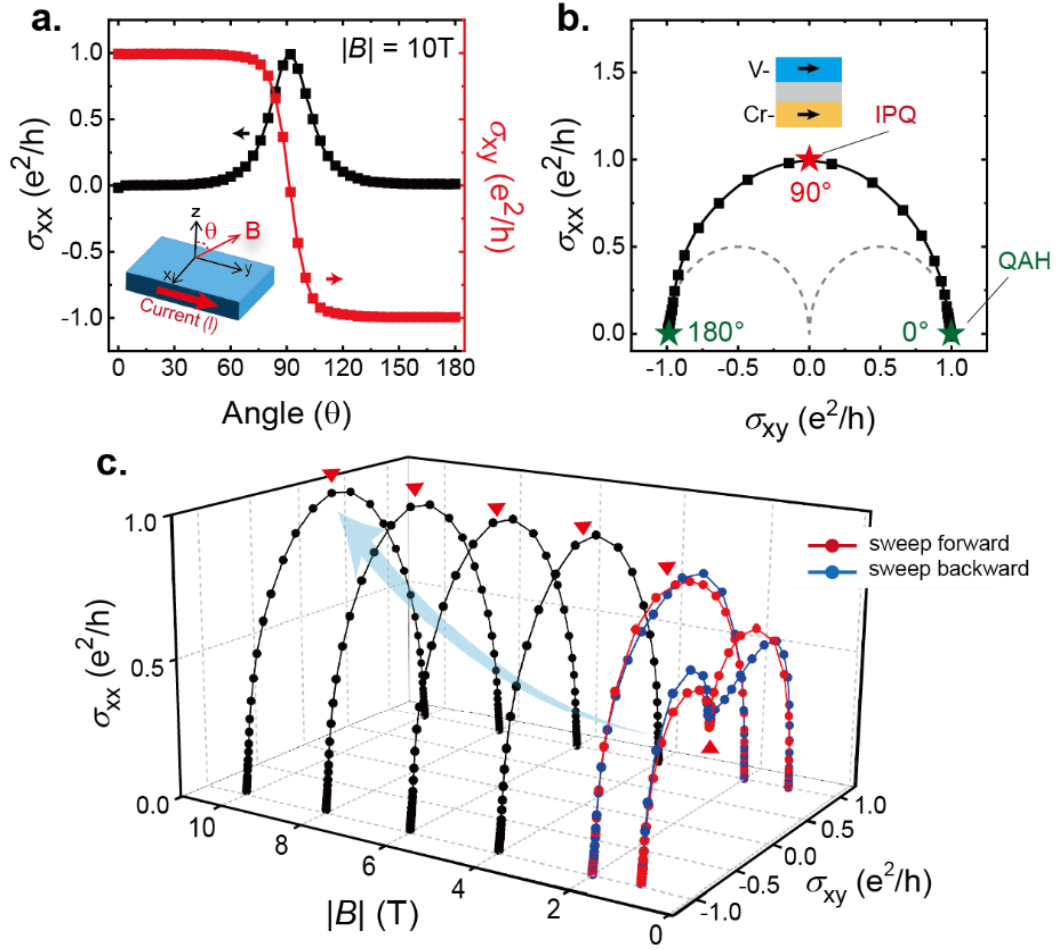

**FIG. 2.** (a) Conductivities $\sigma_{xx}$ and $\sigma_{xy}$ plotted as a function of the angle $\theta$ in the yz plane, with a magnetic field magnitude of $|B| = 10$ T. (b) The corresponding renormalization flow diagram of $\sigma_{xx}$ and $\sigma_{xy}$ where the red and green stars mark the points $(0, e^2/h)$ and the QAH state $(e^2/h, 0)$, respectively. The dashed line indicates flow diagram of ordinary MTIs. (c) The three-dimensional representation of the field angle dependent flow diagram, showing its evolution under different magnetic field strengths.

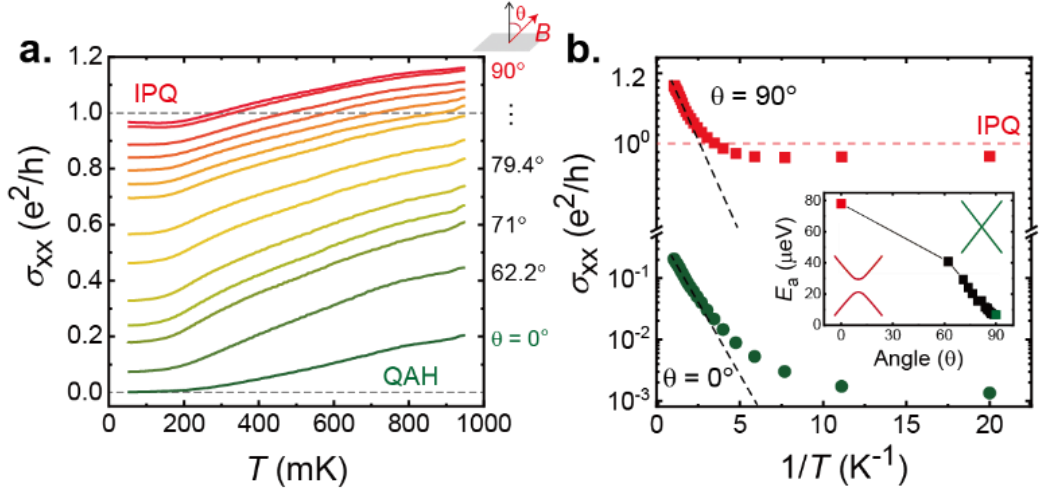

**FIG. 3.** (a) Longitudinal conductivity $\sigma_{xx}$ is shown as a function of temperature, starting from the QAH state. A maple color gradient illustrates the gradual increase in the angle of an 8 T magnetic field relative to the z-axis. (b) A logarithmic-scale plot of $\sigma_{xx}$ is presented as a function of 1/T for two angles: $\theta = 0°$ (green dots) and $\theta = 90°$ (red dots), along with the corresponding fitting lines. The inset shows the activation energy as a function of the magnetic field angle.

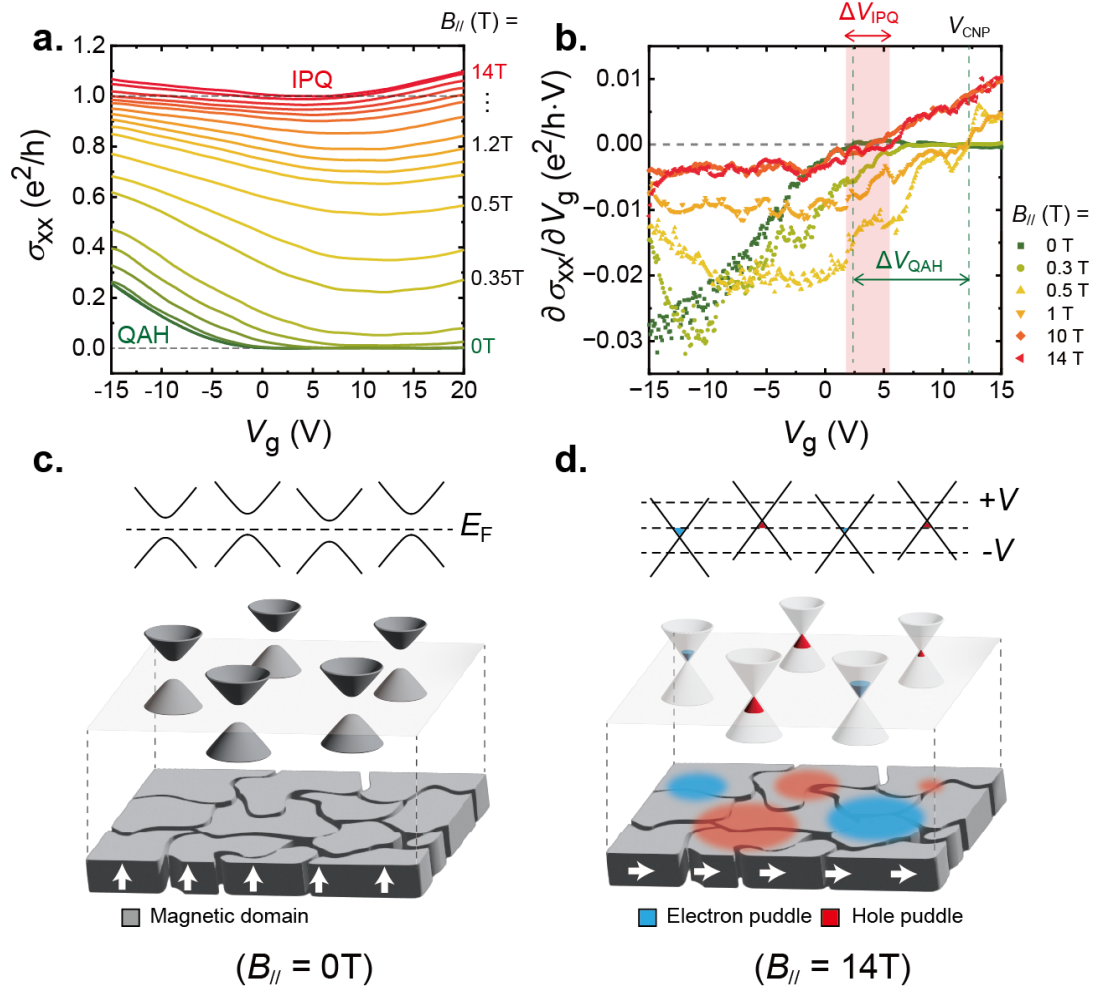

**FIG. 4.** (a) Longitudinal conductivity $\sigma_{xx}$ as functions of gate voltage, starting from the QAH state. A maple color gradient represents the gradual increase in the strength of the in-plane magnetic field. (b) Corresponding derivative of $\sigma_{xx}$ with respect to gate voltage for in-plane fields of 0, 0.3, 0.5, 1, 10, and 14 T, with the gate-voltage ranges for the QAH and IPQ states indicated. (c) Energy landscape of the gapped surface states at zero in-plane field (0 T), where the magnetization of the magnetic domains is saturated along the out-of-plane direction. (d) Energy landscape of the gapless surface states at an in-plane field of 14 T, where the magnetization aligns in the in-plane direction, accompanied by the electron and hole puddles.